\documentclass[twocolumn,groupaddress,aps]{revtex4-2}

\usepackage{graphicx}
\usepackage{dcolumn}
\usepackage{bm}
\usepackage{amsmath}
\usepackage{amssymb}
\usepackage{graphicx}
\usepackage{color} 
\usepackage{ulem}
\usepackage{bm}
\usepackage{braket}
\usepackage[colorlinks,citecolor=blue]{hyperref}
\usepackage{multirow}
\usepackage{makecell}
\usepackage{booktabs}
\usepackage{graphicx}
\usepackage{xcolor}
\usepackage{color}
\usepackage{bbm}

\begin{document}
\title{Perfect transmission of a Dirac particle in one-dimension double square barrier}
\author{Xu Zhang}
\affiliation{Department of Physics and Institute of Theoretical Physics, University of Science and Technology Beijing, Beijing 100083, China}

\author{Qiang Gu}
\email[Corresponding author: ] {qgu@ustb.edu.cn}
\affiliation{Department of Physics and Institute of Theoretical Physics, University of Science and Technology Beijing, Beijing 100083, China}

\begin{abstract}
Dirac particles can undergo perfect transmission through a sufficiently high potential barrier in the Klein zone. Although the perfect Klein tunneling (often referred to as the Klein paradox) is similar to the non-relativistic resonant transmission which occurs only when the kinetic energy exceeds the barrier, the underlying mechanism is believed to be fundamentally distinct. In this work, we show that for the relativistic double-barrier model the perfect-transmission curve can pass continuously from the above-barrier zone to the Klein zone. Additionally, in the Klein zone, perfect transmission occurs even for subcritical barrier heights, supported by both bound-state analysis and wave-packet dynamics. These findings suggest a connection between perfect Klein tunneling and resonant transmission, and provide new insights into the physical nature of the Klein paradox. 
\end{abstract}

\maketitle
\section{Introduction}
\label{sec:level1}
\begin{figure}[t]
\centering
\includegraphics[width=0.46\textwidth]{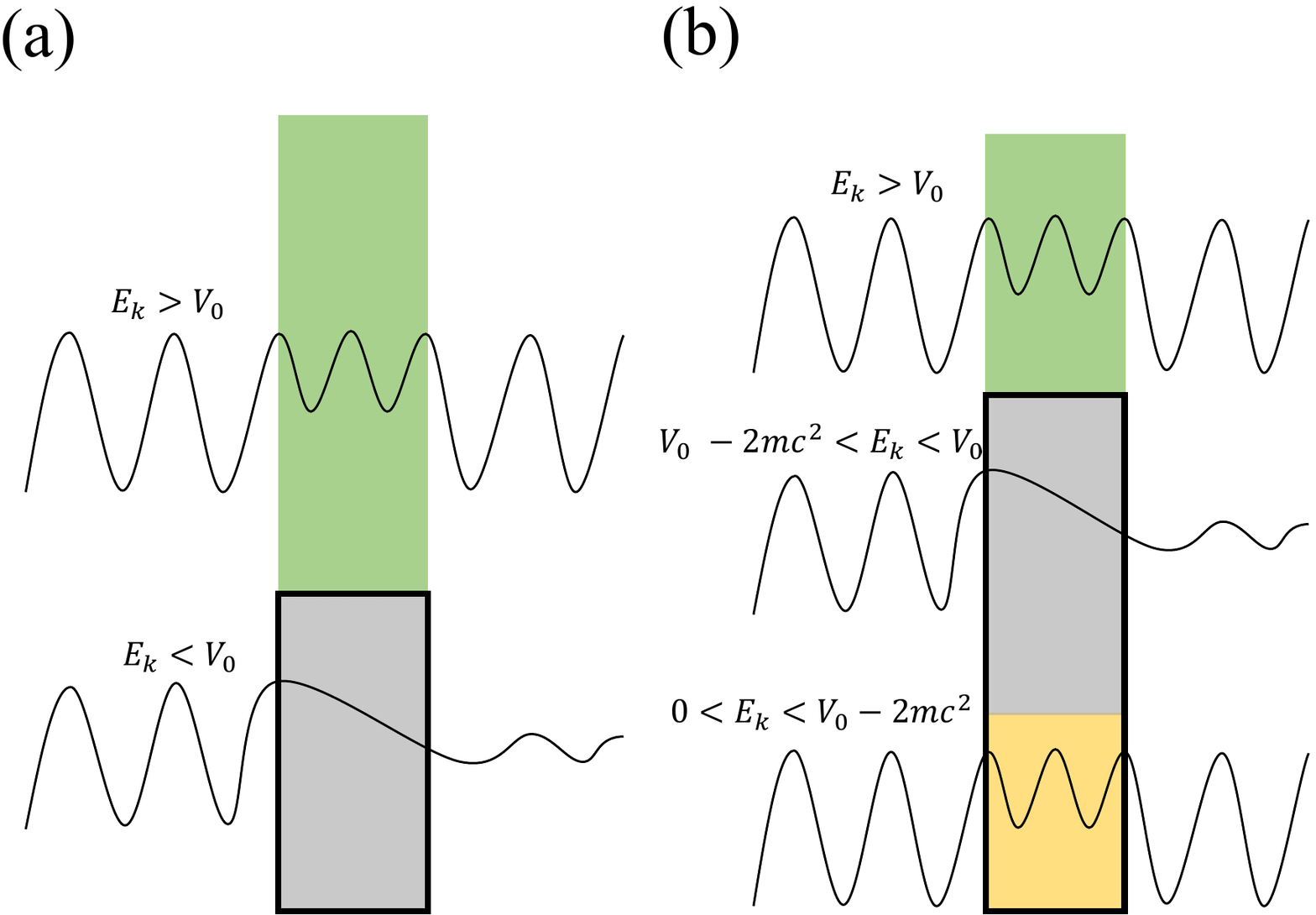}
\caption{Schematic illustration of the particle transmission through a single potential barrier. (a) $V_0 < 2mc^2$ or non-relativistic case. The green shaded region denotes the above-barrier zone where $E_k > V_0$ and resonance transmission with probability reaching 1 may occur. The gray shaded region is the normal-quantum-tunneling zone where $E_k < V_0$ and the transmitted wave is exponentially damped down. (b) $V_0 > 2mc^2$. The Klein zone ($0 < E_k < V_0 - 2mc^2$) appears at the bottom of the barrier, shown as the yellow shaded region.}
\label{barrier}
\end{figure}

Quantum tunneling stands as one of the most celebrated achievements of quantum mechanics, which suggests that a particle can pass through a potential barrier whose height $V_0$ exceeds the kinetic energy of the particle $E_k$ with certain probability \cite{Bohm1951}. On the other hand, although the transmission probability grows with $E_k$ increasing, it can hardly approach 100\%. In the above-barrier energy zone ($E_k > V_0$), perfect transmission with 100\% probability takes place under certain resonance condition, which is called the resonant transmission (RT) \cite{Merzbacher1998}. In relativistic quantum mechanics, the case is similar when the barrier height $V_0 < 2mc^2$, as shown in Fig.~\ref{barrier}(a). However, if $V_0 > 2mc^2$ and the kinetic energy $E_k = E - mc^2$ lies within the energy zone between 0 and $V_0 - 2mc^2$ (named the Klein zone), as shown in Fig.~\ref{barrier}(b), perfect Klein tunneling (PKT) may occur  \cite{Dombey1999}, which is known as the Klein paradox \cite{Klein1929}. Here $E$ and $m$ are the total relativistic energy and the rest mass of the particle, respectively.

Owing to the difficulty in realizing sufficiently high potential barriers, the PKT has not been experimentally confirmed for more than 70 years since its first prediction in 1929. The favorable turn comes with the discovery of graphene \cite{Novoselov2004}, a two-dimensional material whose quasiparticle can be described by a Dirac-like Hamiltonian \cite{Castro}. Since then, PKT is observed in a series of materials with effective Dirac-like particle, such as graphene and Weyl semimetals \cite{Geim2006,Kim2009,Bahat2010,Illes2017,Du2018,Wang2022,Elahi2024}, photonic crystals \cite{Dreisow2012,Zhang2022}, phononic crystals \cite{Jiang2020,Sirota2022,Wu2024}, as well as cold atomic systems \cite{Salger2011,Zhang2012}. These comprehensive achievements have greatly stimulated the renewed interest in this fascinating topic.

Although both the PKT and RT exhibit perfect transmission, it is believed that the underlying mechanism is thoroughly different. Particle scatterings occurring in the above-barrier and the normal-quantum-tunneling zones behave similarly to those of the non-relativistic quantum mechanics, which can be interpreted as the scattering of one particle by the barrier \cite{Greiner}. In contrast, the widely accepted explanation of PKT involves the spontaneous creation of particle-antiparticle pairs \cite{Dombey1999,Su1993,Grobe2004,Matzkin2022}. Nevertheless, we show in this paper that there exists an intriguing connection between the RT in the above-barrier zone and the PKT in the Klein zone, based on a simple double square barrier model.

It is worth noting that the double barrier can exhibit perfect transmission in the normal-quantum-tunneling zone in nonrelativistic quantum mechanics \cite{Hosack1965,Hauge1987}. Physically, this kind of RT has the same origin as that occurring in the above-barrier zone of the single-barrier case, both of which arise from interference of reflected waves from different interfaces. The tunneling of relativistic particles through a double barrier has also attracted considerable interest \cite{Lunardi2007,Bai2007,Villalba2010,Vargas2012}, particularly in the context of condensed-matter physics \cite{Bai2007,Vargas2012}. In this work, we systematically study perfect transmission of Dirac particles across all energy regions. We demonstrate that a perfect-transmission curve can continuously span all regions and provides a link between PKT and RT. This finding provides new insight into the mechanism underlying Klein tunneling.

This work is organized as follows: In Sec.~\ref{sec:level2}, general solutions to the one-dimensional double square potential Dirac equation are derived. In Sec.~\ref{sec:level3}, the transmission of a Dirac particle through a double-square barrier is investigated, and the conditions for perfect transmission in different energy regions are obtained. Sec.~\ref{sec:level4} discusses bound states for the double-square potential wells. In Sec.~\ref{sec:level5}, we study the wave packet dynamics in the presence of a double barrier. The final Sec.~\ref{sec:level6} presents a brief summary.

\begin{figure}
\centering
\includegraphics[width=0.46\textwidth]{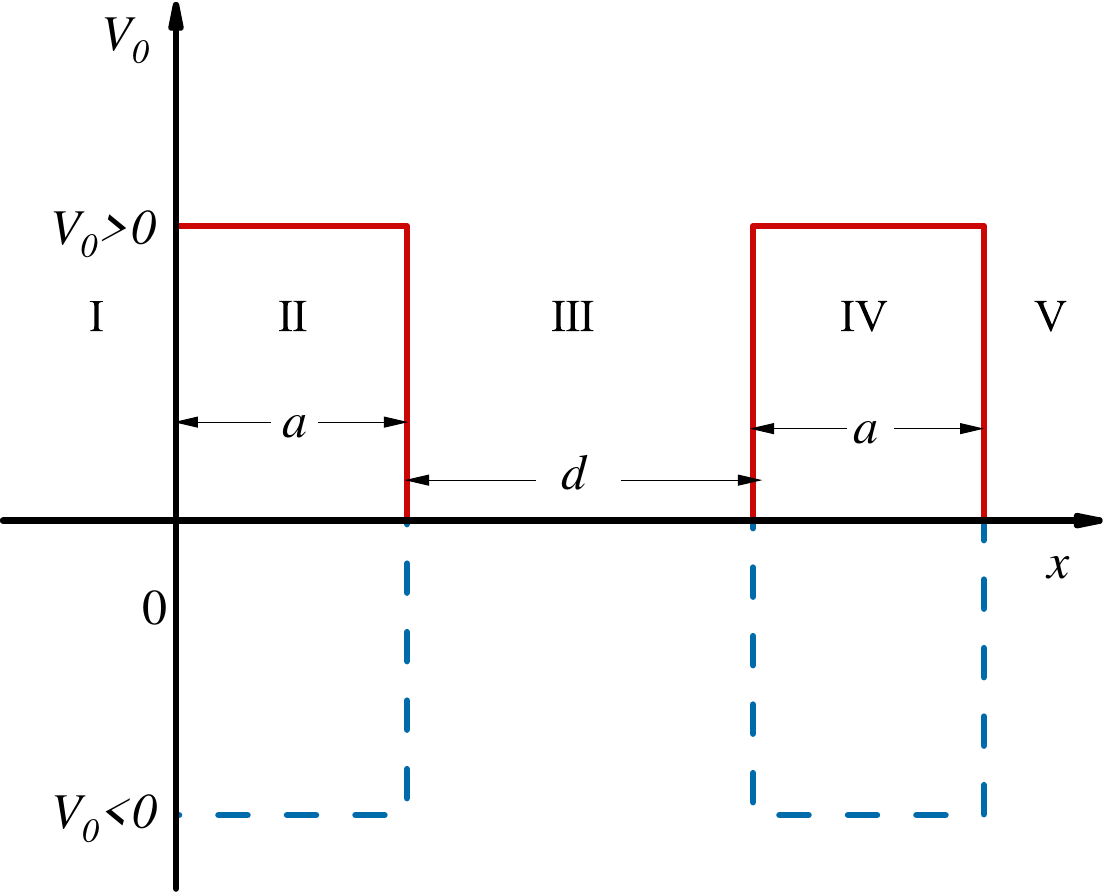}
\caption{Potential barrier (red-solid line) and potential well (blue-dashed line), where $a$ and $d$ denote the barrier width and the distance between the two barriers, respectively.}
\label{model-1}
\end{figure}

\section{\label{sec:level2}The General Solution}To begin with, we choose the representation for $\gamma$ matrices as $\gamma_x$ and $\gamma_0$ to be Pauli matrices $\sigma_x$ and $\sigma_z$ \cite{Dombey2000}, respectively. The one-dimensional Dirac equation with potential $V(x)$ is written as 
\begin{equation}
\left(\hbar c \sigma _x \frac{\partial}{\partial x} -\left( E-V(x) \right)\sigma _z  + mc^2\right)\psi = 0.
\end{equation}
As shown in Fig.~\ref{model-1}, the space is divided into five regions with respect to the potential $V(x)$. $V(x) = 0$ in the free regions (denoted as Regions \uppercase\expandafter{\romannumeral1}, \uppercase\expandafter{\romannumeral3} and \uppercase\expandafter{\romannumeral5}), and $V(x) =V_0$ presents the potential barrier ($V_0>0$) or well ($V_0<0$) regions (Regions \uppercase\expandafter{\romannumeral2} and \uppercase\expandafter{\romannumeral4}). Then the wave functions in the free regions $\psi_{\alpha}$ ($\alpha=1, 2, 3$) and potential regions $\psi_{\beta}$ ($\beta=1, 2$) are given by
\begin{subequations}\label{wavefunct}
\begin{align}
\psi_{\alpha} =& A_{\alpha} 
\begin{pmatrix}
1\\ \frac{-ikc}{E+mc^2}
\end{pmatrix}
e^{i\frac{kx}{\hbar}}+B_{\alpha}
\begin{pmatrix}
1 \\ \frac{ikc}{E+mc^2}
\end{pmatrix}
e^{-i\frac{kx}{\hbar}},
\\
\psi_{\beta} =& C_{\beta} 
\begin{pmatrix}
1 \\ \frac{-ipc}{E-V_0+mc^2}
\end{pmatrix}
e^{i\frac{px}{\hbar}}\nonumber \\ &+D_{\beta} 
\begin{pmatrix}
1\\ \frac{ipc}{E-V_0+mc^2}
\end{pmatrix}
e^{-i\frac{px}{\hbar}},
\end{align}
\end{subequations}
where $k$ and $p$ represent the momentum of particles in the free region and potential region, respectively. They can be calculated according to following equations,
\begin{subequations}\label{momentum}
\begin{align}
	k^2c^2 &= E^2 -m^2c^4 ,	\\
	p^2c^2 &= (E - V_0)^2 -m^2c^4 .
\end{align} 
\end{subequations}
It can be seen that, depending on $E$ and $V_0$, $k$ and $p$ may be real or imaginary. $k$ is real outside the barrier (free propagation) and imaginary otherwise (no propagation), while $p$ is real inside the barrier for RT (or PKT) and imaginary otherwise (exponential decay).

\begin{table*}[t]
\begin{center}
\caption{Characteristics of momentum $k$ and $p$ in the potential barrier model: $V_0 >0$}
\label{scatteringtable}
\begin{ruledtabular}
\begin{tabular}{c c c c c c c c}
\multicolumn{4}{c}{$V_0<2mc^2$} & \multicolumn{4}{c}{$V_0>2mc^2$}\\
\hline
Incident energy &k &p &$\gamma$ &Incident energy &k &p &$\gamma$\\
\hline
$V_0<E_k$ &real &real &real &$V_0<E_k$ &real &real &real\\

\multirow{2}{*}{$0<E_k<V_0$} &\multirow{2}{*}{real} &\multirow{2}{*}{imaginary} &\multirow{2}{*}{imaginary} &$V_0-2mc^2<E_k<V_0$ &real  &imaginary  &imaginary \\ &  &  & &$0<E_k<V_0-2mc^2 $ &real &real  &real
\end{tabular}
\end{ruledtabular}
\end{center}
\end{table*}

The coefficients in Eqs.~(\ref{wavefunct}) can be determined by the continuity condition that the wave function should be continuous at region boundaries. Regarding Regions \uppercase\expandafter{\romannumeral1} and \uppercase\expandafter{\romannumeral3}, we obtain
\begin{equation}\label{firstbarrier}
\begin{pmatrix}
A_1\\B_1
\end{pmatrix}
=
\begin{pmatrix}
m_1 & m_2 \\ m_3 & m_4
\end{pmatrix}
\begin{pmatrix}
A_2\\B_2
\end{pmatrix},
\end{equation}
where
\begin{subequations}\label{m}
\begin{align}
m_1 =&\frac{1}{4}\left[\left(2+\gamma + \frac{1}{\gamma}\right)e^{i\frac{ka - pa}{\hbar}}\nonumber\right.\\&\left. + \left(2 - \gamma -\frac{1}{\gamma}\right)e^{i\frac{ka + pa}{\hbar}}\right],\\
m_2 =&\frac{1}{4}\left[\left(-\gamma + \frac{1}{\gamma}\right)e^{-i\frac{ka + pa}{\hbar}} \nonumber\right.\\&\left.+ \left(\gamma - \frac{1}{\gamma}\right)e^{i\frac{-ka + pa}{\hbar}}\right],\\
m_3 =&\frac{1}{4}\left[\left(\gamma - \frac{1}{\gamma}\right)e^{i\frac{ka - pa}{\hbar}} \nonumber\right.\\&\left.+ \left(-\gamma + \frac{1}{\gamma}\right)e^{i\frac{ka + pa}{\hbar}} \right],\\
m_4 =&\frac{1}{4}\left[\left(2-\gamma - \frac{1}{\gamma}\right)e^{-i\frac{ka + pa}{\hbar}} \nonumber\right.\\&\left.+ \left(2 + \gamma +\frac{1}{\gamma}\right)e^{i\frac{-ka + pa}{\hbar}} \right],
\end{align}
\end{subequations}
with
\begin{equation}\label{gamma}
\gamma = \frac{E+mc^2-V_0}{E+mc^2} \frac{k}{p}.
\end{equation}
And the relation between Regions \uppercase\expandafter{\romannumeral3} and \uppercase\expandafter{\romannumeral5} is given by 
\begin{equation}\label{secondbarrier}
\begin{pmatrix}
A_2\\B_2
\end{pmatrix}
= 
\begin{pmatrix}
m_1 & m_2 e^{-i\frac{k(2d + 2a)}{\hbar}} \\ m_3 e^{i\frac{k(2d + 2a)}{\hbar}} & m_4
\end{pmatrix}
\
\begin{pmatrix}
A_3\\B_3
\end{pmatrix}.
\end{equation}

By combining Eqs. (\ref{firstbarrier}) and (\ref{secondbarrier}), the relation between Regions \uppercase\expandafter{\romannumeral1} and \uppercase\expandafter{\romannumeral5} can be obtained as 
\begin{equation}\label{TMatrix}
\begin{pmatrix}
A_1\\B_1
\end{pmatrix}
= 
\begin{pmatrix}
M_1 & M_2 \\ M_3 & M_4
\end{pmatrix}
\begin{pmatrix}
A_3 \\ B_3
\end{pmatrix},
\end{equation}
with
\begin{subequations}\label{M}
\begin{align}
M_1 &=m_1^2 + m_2 m_3 e^{i\frac{k(2d + 2a)}{\hbar}},\\
M_2 &=m_1 m_2 e^{-i\frac{k(2d + 2a)}{\hbar}} + m_2 m_4,\\
M_3 &=m_1 m_3 + m_3  m_4  e^{i\frac{k(2d + 2a)}{\hbar}},\\
M_4 &=m_2  m_3 e^{-i\frac{k(2d + 2a)}{\hbar}}+m_4^2.
\end{align}
\end{subequations}

Above solutions are generally valid for the one-dimensional Dirac equation regardless of $V_0 \ge 0$ or $V_0 < 0$. For the barrier case, there exist only scattering states. On the other hand, bound states may appear in the well. For clarity, we define the kinetic energy of the particle as $E_k=E-mc^2$, which can be positive or negative, corresponding to scattering states ($E_k>0$) or bound states ($E_k<0$), respectively.

\section{\label{sec:level3}Scattering states} 
Now we discuss the scattering states that have the energy $E_k>0$, regardless of $V_0>0$ or $V_0<0$. Supposing that the incident particle is from left to right, and the wavefunction is describing by the $A_1$ term in Eq.~(\ref{wavefunct}a). Then the $B_1$ ans $A_3$ terms describes reflected ans outgoing part, respectively. And it is reasonable to set the parameter $B_3=0$. Then the transmission coefficient is given by
\begin{equation}
	T =\left|\frac{A_3}{A_1}\right|^2 =\left|\frac{1}{M_1}\right|^2 ,
	\label{T}
\end{equation}
with the parameters are determined from Eqs. (\ref{m}) and Eq.~(\ref{M}a) as 
\begin{equation}\label{M_1}
\begin{aligned}
M_1 =& \frac{1}{16}e^{2i\frac{ka}{\hbar}}\left(e^{i\frac{pa}{\hbar}}+e^{-i\frac{pa}{\hbar}}\right)^2 \\ &  \times \left\{ \left[\left(\gamma + \frac{1}{\gamma}\right)^2 -\left(\gamma - \frac{1}{\gamma}\right)^2 e^{2ikd} \tanh^2\!\left(\frac{ipa}{\hbar}\right) \right]\right.\\& \left.-4\left(\gamma+\frac{1}{\gamma}\right)\tanh\!\left(\frac{ipa}{\hbar}\right)+4 \right\}.
\end{aligned} 
\end{equation}

Since $E_k>0$ for scattering states, $k$ is always real. Therefore, the scattering behavior mainly depends on whether $p$ is real or imaginary. Table \ref{scatteringtable} summarizes whether $k$ , $p$ and $\gamma$ take real or imaginary values for different combinations of the barrier height and the incident particle energy.
\begin{figure*}[t]
	\centering
	\includegraphics[width=0.96\textwidth]{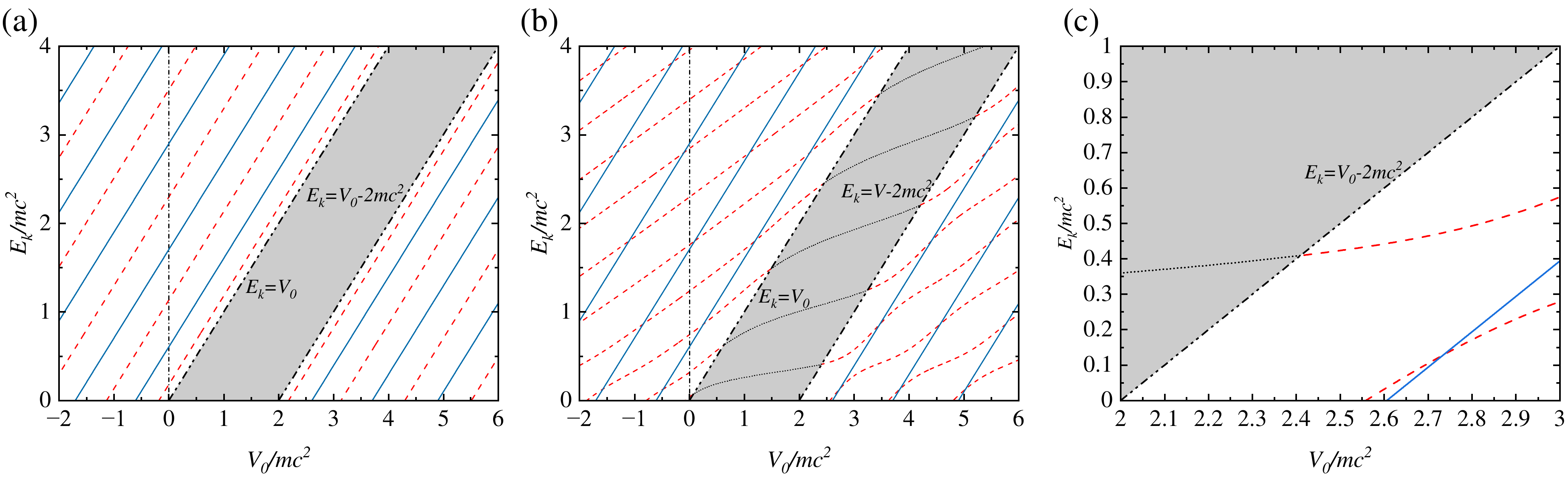}
	\caption{Perfect transmission curves for Dirac particles through a one-dimensional double potential barrier, each barrier of width $a =2.5\lambda$. (a) and (b) correspond to inter-barrier distances $d = 0\lambda$ and $d = 5\lambda$, respectively. (c) Close-up of a part in (b). The above-barrier zone lies above the black dash-dot-dot line $E_k =V_0$ and the Klein zone is below the black dash-dot-dot line $E_k=V_0-2mc^2$. The gray region between the two lines ($E_k =V_0$ and $E_k=V_0-2mc^2$) denotes the normal-tunneling zone in which the transmission probability decays exponentially as the barrier height and width increase. The blue-solid, red-dashed, and black-dotted curves represent perfect-transmission conditions. The intersections of the perfect-transmission curves with the horizontal axis ($E_k = 0$) correspond respectively to the zero-energy resonance ($V_0<0$) and the zero-momentum resonance ($V_0>2mc^2$).}
\label{perfect tunneling}
\end{figure*}
\subsection{The real $p$ case} 
 A real $p$ corresponds to the propagating wave that allows for perfect transmission (the above-barrier and the Klein zones). When $E_k>V_0$, $p$ is always real and subsequently $\gamma$ is real according to Eq.~(\ref{gamma}), as denoted in Table \ref{scatteringtable}. Note that this condition is naturally satisfied for the potential well case in which $V_0<0$ in Regions  \uppercase\expandafter{\romannumeral2} and  \uppercase\expandafter{\romannumeral4}. By substituting Eq.~(\ref{M_1}) into Eq.~(\ref{T}), we obtain the transmission coefficient
\begin{equation}
\begin{aligned}
T =\Biggl\{1 &+ \left(\gamma - \frac{1}{\gamma}\right)^2 \sin^2\!\left(\frac{pa}{\hbar}\right) \\ & \times \Biggl[ \frac{1}{8}\left(\gamma + \frac{1}{\gamma}\right)^2 \sin^2\!\left(\frac{pa}{\hbar}\right)\left[1 - \cos\!\left(\frac{2kd}{\hbar}\right)\right]\\ & + \frac{1}{2} \cos^2\!\left(\frac{pa}{\hbar}\right) \left[\cos\!\left(\frac{2kd}{\hbar}\right) + 1\right] \\ &  -\frac{1}{2}\left(\gamma + \frac{1}{\gamma}\right)\sin\!\left(\frac{pa}{\hbar}\right)\cos\!\left(\frac{pa}{\hbar}\right)\sin\!\left(\frac{kd}{\hbar}\right)\Biggl]\Biggl\}^{- 1}  \label{cofficient1}.
\end{aligned}
\end{equation}
Then we can obtain the conditions for perfect transmission ($T = 1$) as follows
\begin{subequations}\label{resonence1}
\begin{align}
\sin\!\left(\frac{pa}{\hbar}\right) &= 0, \\ \text{or} \quad  \cot\!\left(\frac{pa}{\hbar}\right) &= \frac{1}{2} \left(\gamma + \frac{1}{\gamma}\right)\tan\!\left(\frac{kd}{\hbar}\right).
\end{align}
\end{subequations}
Note that the above conditions hold both in the above-barrier zone and in the Klein zone. By substituting Eqs. (\ref{momentum}) and (\ref{gamma}) into Eqs. (\ref{resonence1}), it's easy to get that Eqs. (\ref{resonence1}) characterize the relation between $E_k$ and $V_0$, as shown in Fig.~\ref{perfect tunneling}. Eqs. (\ref{resonence1}a) and (\ref{resonence1}b) denote the odd-parity and even-parity condition, shown as the blue-solid and red-dashed curves, respectively. In particular, when the rest mass $m=0$, we have $\gamma = \frac{E+0-V}{E+0}\frac{E}{E-V} = 1$, which directly gives $T=1$ for Eq.~(\ref{cofficient1}). This result is consistent with the phenomenon of perfect transmission for normally incident massless fermions in graphene \cite{Geim2006,Vargas2012,Bai2007}. In the following, we focus on the case of massive particles.

When the barrier distance $d=0$, the double-barrier merges into a single barrier with double width $2a$. Then Eqs. (\ref{resonence1}) reduces to $\sin\!\left(\frac{pa}{\hbar}\right)=0$ (odd-parity condition) and $\cos\!\left(\frac{pa}{\hbar}\right)=0$ (even-parity condition), which is the perfect-transmission condition through a single barrier of width $2a$ \cite{Greiner}. Eqs. (\ref{resonence1}) simplified to the following $E_k-V_0$ relationship:
 \begin{equation}
 E_k = V_0 -mc^2 \pm mc^2 \sqrt{1 + \frac{n^2\lambda^2}{16a^2}},
 \label{EVrelation2a}
 \end{equation}
where $\lambda=\frac{h}{mc}$ is the Compton wavelength and $n$ is an integer. The above equation with plus or minus sign inside represents the perfect transmission for the above-barrier zone or the Klein zone, respectively. This equation also clearly demonstrates the linear relation between $E_k$ and $V_0$, which accounts for the observation that all the perfect-transmission curves in Fig.~\ref{perfect tunneling}(a) are parallel.
\begin{table*}[t]
\begin{center}
\caption{Characteristics of momentum $k$ and $p$ in the potential well model: $V_0<0$}
\label{boundtable}
\begin{ruledtabular}
\begin{tabular}{c c c c c c c c}
\multicolumn{4}{c}{$V_0>-2mc^2$} & \multicolumn{4}{c}{$V_0<-2mc^2$}\\
\hline
Incident kinetic energy &k &p &$\gamma$ &Incident kinetic energy &k &p &$\gamma$\\
\hline
$0<E_k$ &real &real &real &$0<E_k$ &real &real &real\\

$V_0<E_k<0$ &imaginary &real &imaginary &$-2mc^2<E_k<0$ &imaginary &real &imaginary\\

$-2mc^2<E_k<V_0$ &imaginary &imaginary &real &$V_0<E_k<-2mc^2$ &real  &real  &real \\

 $V_0-2mc^2<E_k<-2mc^2 $ &real  &imaginary  &imaginary &$V_0-2mc^2<E_k<V_0 $ &real &imaginary  &imaginary·\\

 $E_k<V_0 -2mc^2 $ &real  &real  &real &$E_k<V_0 -2mc^2 $ &real &real  &real
\end{tabular}
\end{ruledtabular}
\end{center}
\end{table*}

The effect of the barrier separation $d$ on Klein tunneling has also attracted interest, for example through numerical studies \cite{Villalba2010}. Here, we present a systematic analysis on this issue. As the single-barrier splits ($d\ne 0$), the odd-parity condition remains while the even-parity condition changes. Comparing Fig.~\ref{perfect tunneling}a with Fig.~\ref{perfect tunneling}b, all the blue-dashed curves keep unaffected, whereas the red-dashed curves are tilted rightward due to the finite $d$.  And larger $d$ results in stronger tilt. The red-dashed curves in the above-barrier zone (RT) terminate at the boundary of the normal quantum-tunneling zone. On the other hand, additional perfect-transmission curves (PKT) emerge in the Klein zone.

It is worth noting that at zero kinetic energy ($E_k=0$), perfect transmission in the above-barrier zone is referred to as the zero-energy resonance, while that in the Klein zone is termed the zero-momentum resonance \cite{Dombey2000, Dombey2002}. The zero-energy resonance occurs when low-energy particles perfectly transmit through a potential well supporting a half-bound state. Similarly, the zero-momentum resonance corresponds to perfect transmission of low-momentum particles scattered by a potential barrier whose strength supports a supercritical state of antiparticles \cite{Dombey2000}. The lowest intersection of the red-dashed curves with the horizontal axis in the Klein zone is the supercritical threshold. Above this threshold, antiparticle bound states inside the barrier can resonate with the particle continuum outside, giving rise to PKT. It is well-established in the single-barrier case that all PKT curves appear to the right side of the first zero-momentum resonance point (the supercritical threshold), as clearly seen in Fig.~\ref{perfect tunneling}(a).

The zero-momentum resonance points can be calculated according to Eqs.~(\ref{resonence1}) by setting $E_k=0$. Naturally, those resonance points   determined by Eq.~(\ref{resonence1}a) are independent of the separation $d$, while those determined by Eq.~(\ref{resonence1}b) depend on $d$. At $E_k=0$, Eq.~(\ref{resonence1}b) reduces to:
\begin{subequations}\label{zeroenergy}
\begin{align}
\cot\!\left(\frac{pa}{\hbar}\right) &= \frac{mc^2d\sqrt{-V_0}}{\sqrt{-V_0 + 2mc^2}}, \quad V_0<0, \\
\cot\!\left(\frac{pa}{\hbar}\right) &= \frac{mc^2d\sqrt{V_0}}{\sqrt{V_0-2mc^2}}, \quad V_0>0,
\end{align}
\end{subequations}
in which the parameter $d$ appears. As is clearly seen from Fig.~\ref{perfect tunneling}(b), both the zero-energy resonance and the zero-momentum resonance are affected solely by the distance $d$ through the perfect-transmission curves associated with the even-parity condition (red-dashed curves). Furthermore, as the distance $d$ increases, the minimum potential well depth required for zero-energy resonance decreases, while the minimum barrier height required for zero momentum resonance increases. A more detailed discussion of the bound-state problem related to zero-energy and zero-momentum resonances will be given in Sec. \ref{sec:level4}.
\begin{figure*}[t]
\centering
\includegraphics[width=0.96\textwidth]{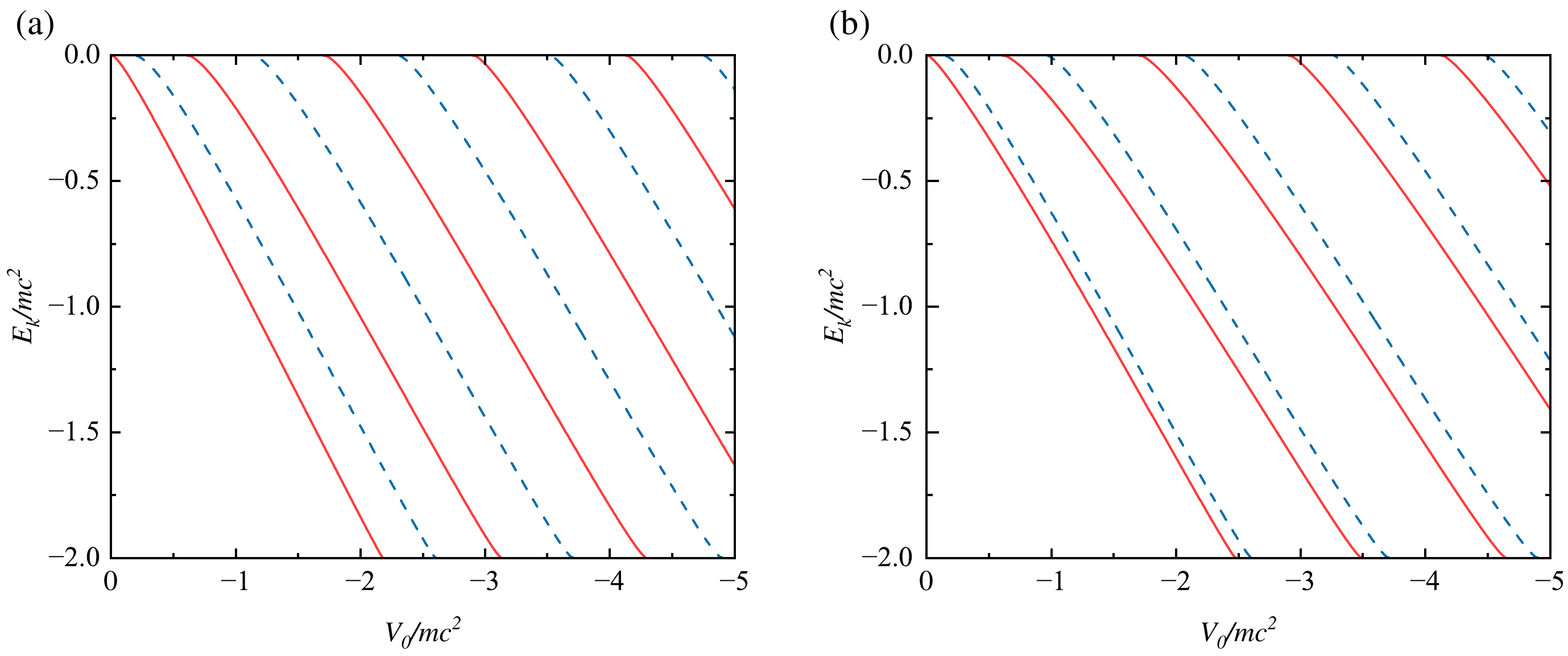}
\caption{Eigenvalue spectrum of bound Dirac particles in a one-dimensional double potential well, each well of width $a = 2.5\lambda$. (a) and (b) correspond to inter-well distances $d = 0\lambda$ and $d = 1\lambda$, respectively. The intersections of the curve with the upper and lower axes correspond respectively to the half-bound and supercritical states, which are related to the zero-energy and zero-momentum resonances.}
\label{eigenvalue spectrum}
\end{figure*}
\subsection{The imaginary $p$ case} 
 When the momentum $p$ is imaginary, the wave function exhibits exponential decay (evanescent wave) inside the barrier region (the normal-quantum-tunneling zone). That is, the gray region in Fig.~\ref{perfect tunneling} separates the Klein zone from the above-barrier zone. For the case where $p$ and $\gamma$ are imaginary, we set $\gamma = i\delta$ and $p = iq$ and substituting them into Eq.~(\ref{M_1}), then inserting the resulting expression into Eq.~(\ref{T}), we obtain
\begin{equation}\label{cofficient2}
\begin{aligned}
T = \Biggl\{ 1  +& \frac{1}{16}\left(\delta + \frac{1}{\delta}\right)^2 \left(e^{\frac{qa}{\hbar}} - e^{-\frac{qa}{\hbar}}\right)^2  \\ & \times \Biggl[ \left(e^{\frac{qa}{\hbar}} + e^{-\frac{qa}{\hbar}}\right)^2  \\& + \frac{1}{4} \left(\delta + \frac{1}{\delta}\right)^2 \left(e^{\frac{qa}{\hbar}} - e^{-\frac{qa}{\hbar}}\right)^2 \sin^2\!\left(\frac{kd}{\hbar}\right) \\&  + \left(\delta - \frac{1}{\delta}\right)\left(e^{\frac{2qa}{\hbar}} - e^{-\frac{2qa}{\hbar}}\right)\sin\!\left(\frac{kd}{\hbar}\right)\cos\!\left(\frac{kd}{\hbar}\right) \\& - 2\left(e^{\frac{2qa}{\hbar}} + e^{-\frac{2qa}{\hbar}}\right)\sin^2\!\left(\frac{kd}{\hbar}\right)\Biggl] \Biggl\}^{-1}.
\end{aligned}
\end{equation}
It can be seen that the prefactor in the second term on the right-hand side of the equality is the same as in the single-barrier case, and it grows exponentially with increasing barrier height and width. When $d=0$, it is clear that $T<1$, implying the absence of perfect transmission. In the double-barrier case ($d \neq 0$), perfect transmission can occur provided that
\begin{equation}\label{resonence2}
	\coth\!\left(\frac{qa}{\hbar}\right) = \frac{1}{2}\left(\frac{1}{\delta} - \delta\right) \tan\!\left(\frac{kd}{\hbar}\right).
\end{equation}
This result indicates that the presence of a finite distance $d$ gives rise to perfect transmission within the normal-quantum-tunneling zone. In the non-relativistic limit, the rest energy $mc^2$ is much greater than the kinetic and potential energy. According to Eqs. (\ref{momentum}), we have $k = \sqrt{2mE_k}$, $p = \sqrt{2m(E_k-V)}$ and $\gamma=k/p$. By substituting $\gamma = i\delta$ and $p = iq$  into Eq.~(\ref{resonence1}), we obtain a perfect-transmission condition identical to that derived from the Schr\"{o}dinger equation for a symmetric double-barrier system in the normal-quantum-tunneling zone \cite{Hauge1987}.
 
As shown in Fig.~\ref{perfect tunneling}(b), it is evident that the perfect-transmission curve in the normal-quantum-tunneling zone continuously connects to those in the Klein and above-barrier zones. Furthermore, it can be proven that the connection points are continuous. The fact that the perfect-transmission curve spans continuously from the above-barrier zone to the Klein zone suggests that perfect transmission in all three zones may share a similar underlying mechanism. If so, perfect transmission in the Klein zone does not necessarily require spontaneous particle–antiparticle pair creation, since perfect transmission in the above-barrier zone certainly does not.
Further insight can be gained from Fig.~\ref{perfect tunneling}(c), where a segment of the perfect-transmission curve lies in the Klein zone but below the supercritical threshold and particle–antiparticle pair creation is not expected to occur in this energy region. This indicates that, in a double-barrier system, PKT can occur without an apparent involvement of spontaneous particle–antiparticle pair creation. By contrast, in the single-barrier case, PKT occurs only when the barrier exceeds the supercritical threshold.

\section{\label{sec:level4}Bound states} 
We now turn to the discussion of bound states. For the bound state, the wave function is localized in the well ($V_0<0$) and it should tend to zero as $x\to -\infty$ in Region \uppercase\expandafter{\romannumeral1} and as $x\to \infty$ in Region \uppercase\expandafter{\romannumeral5}. To meet this requirement, the momentum $p$ should be real but $k$ is imaginary, i.e., $k=i\kappa$ ($\kappa$ is real). Meanwhile, the parameter $\gamma = i\delta$.

As Table \ref{boundtable} shows, bound states exist in the energy region $V_0<E_k<0$ when $-2mc^2<V_0<0$, and in the energy region $V_0-2mc^2<E_k<0$ when $V_0<-2mc^2$. It is convenient to set $A_1$ and $B_3$ to be $0$. Meanwhile, $B_1$ and $A_3$ cannot be zero to ensure that the wave function does not vanish. Thus $M_1=0$ according to Eq.~(\ref{M_1}), we have 
\begin{equation}
\cot\!\left(\frac{pa}{\hbar}\right)=-\frac{1}{2}\left[\left(\delta-\frac{1}{\delta}\right)\pm \left(\delta+\frac{1}{\delta}\right)e^{-\frac{\kappa d}{\hbar}}\right].
\label{bound-solu}
\end{equation} 
Substituting Eqs.~ (\ref{momentum}) and (\ref{gamma}) into Eq.~(\ref{bound-solu}) (which represents the $E_k-V_0$ relationship) allows us to obtain the eigenvalue spectrum when bound states exist. As shown in Fig.~\ref{eigenvalue spectrum}, the blue-dashed line corresponds to the plus branch in Eq.~(\ref{bound-solu}), and the red-solid line corresponds to the minus branch.

When $d \to \infty$, the double-well can be regarded as two separate single wells and Eq.~(\ref{bound-solu}) reduces to $\cot\!\left(\frac{pa}{\hbar}\right)=-\frac{1}{2}\left(\delta-\frac{1}{\delta}\right)$.
It is just the solution for the bound state in a single well with the well-width $a$ \cite{Greiner}. When two wells come close to each other, two states in each well with same energy originally undergo splitting due to the tunneling effect. At $d=0$, the two wells merge into one well with the well-width $2a$, and Eq.~(\ref{bound-solu}) reduces to $\cot\!\left(\frac{2pa}{\hbar}\right)=-\frac{1}{2}\left(\delta-\frac{1}{\delta}\right)$. 

\begin{figure*}[t]
\centering
\includegraphics[width=0.96\textwidth]{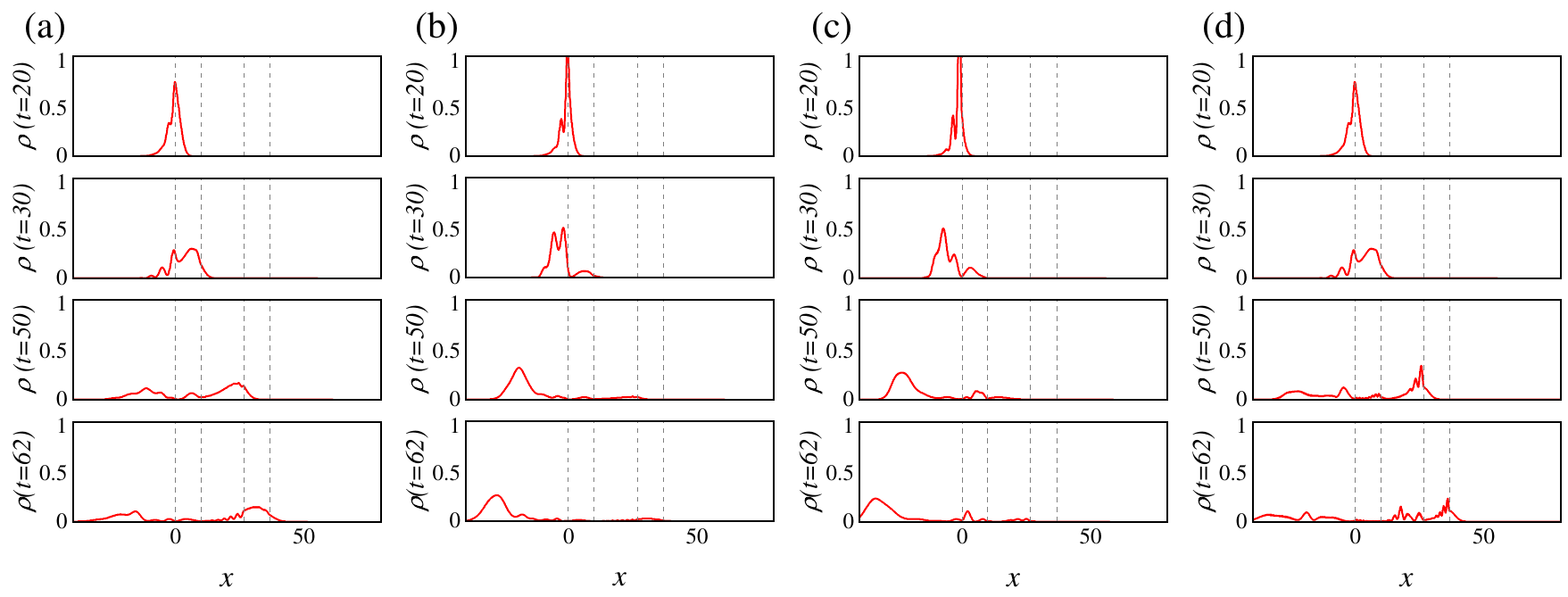}
\caption{Wave-packet dynamics of a Dirac particle impinging on a double barrier. The barrier heights in panels (a)–(d) are $V=0.4$, $V=0.8$, $V=2.55$, and $V=4$, respectively. The wave packet is represented by the red solid curve. Two separated potential barriers are enclosed by the gray dashed lines. In all panels, the same initial state is used, and the barriers are identical except for their heights. Panels (a)–(d), from top to bottom, show the time evolution of the wave packet in the above-barrier zone, the normal-quantum-tunneling zone, the Klein zone below the supercritical threshold, and the Klein zone above the supercritical threshold, respectively. The parameters used in the simulations are given in the main text. }
\label{wavepackets}
\end{figure*}

For the half-bound state, as $E_k \to 0$ one has $\kappa \to 0$ and $\delta\to \infty$ (see Eqs.~(\ref{momentum}) and (\ref{gamma})), which lead to the following relations
\begin{subequations}
\begin{align}
\cot\!\left(\frac{pa}{\hbar}\right) &= \frac{mc^2d\sqrt{-V_0}}{\sqrt{-V_0 + 2mc^2}}, \\ \text{or} \quad
\sin\!\left(\frac{pa}{\hbar}\right) &= 0,
\end{align}
\label{halfboundstate}
\end{subequations}
In a similar manner, the supercritical states at $E_k=-2mc^2$ are
\begin{subequations}
\begin{align}
\sin\!\left(\frac{pa}{\hbar}\right) &= 0 , \\ \text{or} \quad
\cot\!\left(\frac{pa}{\hbar}\right) &= \frac{mc^2d\sqrt{-V_0}}{\sqrt{-V_0-2mc^2}}.
\end{align}
\label{supercriticalstates}
\end{subequations}
It can be clearly seen that Eqs.~(\ref{halfboundstate}a) and (\ref{supercriticalstates}b) are identical to Eq.~(\ref{zeroenergy}a) and Eq.~(\ref{zeroenergy}b), respectively. And Eqs.~(\ref{supercriticalstates}a) and (\ref{halfboundstate}b) are both coincide with Eq.~(\ref{resonence1}a). As shown in Figs.~\ref{perfect tunneling} and \ref{eigenvalue spectrum}, the half-bound and supercritical states occur at potential values corresponding to the zero-energy and zero-momentum resonances, respectively. Compared with the half-bound state case, the supercritical state is affected differently. For the half-bound state, the plus branch is influenced, while for the supercritical state the minus branch is influenced. Nevertheless, in both cases only the even-parity branch is modified by the distance $d$. As shown in Fig.~\ref{eigenvalue spectrum}b, increasing the separation $d$ requires a deeper potential well for the first supercritical state to appear. This behavior is consistent with the scattering-state result that a larger $d$ increases the barrier height required for the first zero-momentum resonance.

\section{\label{sec:level5}Wave-packet dynamics}
The stationary scattering analysis presented above identifies resonance structures in the three energy regions considered. In order to further characterize the feature of tunneling in each region, we investigate the time-dependent propagation of wave packets across the barriers in this section. Without loss of generality, the initial state is chosen to be predominantly composed of positive-energy components, given by
\begin{equation}
\Psi(0, x) = \begin{pmatrix} 1 \\ \frac{-ik_0}{E_{0} + m} \end{pmatrix} e^ { -\frac{(x - x_0)^2}{2\sigma^2} + i k_0 x }.
\end{equation}
Natural units with $\hbar=c=1$ are used in this section. The initial wave packet is centered at $x_0$, which is located far to the left of the potential barrier. By choosing $k_0>0$, the wave packet propagates from left to right. The time evolution is obtained numerically by applying a finite-difference approximation to the Hamiltonian. According to our choice of representation, the Hamiltonian takes the form
\begin{equation}
H=-\sigma_y p + V(x) +\sigma_z m.
\end{equation}
The corresponding time-evolution operator is then given by
\begin{equation}
U(\delta t) = e^{-iH\delta t} = \begin{pmatrix} e^{-im\delta t} & 0 \\ 0 &  e^{-im\delta t} \end{pmatrix} \sum_{n=0}^{n_{max}} \frac{1}{n!}u^2 ,
\end{equation}
where $u = -i\delta t\begin{pmatrix} V(x) & \partial_x \\ -\partial_x & V(x) \end{pmatrix}$. 
The time-evolution operator is constructed using a fourth-order finite-difference approximation for the spatial derivatives. For convenience, we adopt a quasi-square potential barrier with smooth boundaries following previous studies \cite{KennedyW,Alkhateeb2021}, which is given by
\begin{equation}\label{smooth barrier}
\begin{aligned}
V(x) = \frac{V_0}{2}( &\tanh[\epsilon x] -\tanh[\epsilon (x-a)] \\&+\tanh[\epsilon (x-a-d)]-\tanh[\epsilon (x-2a-d)]).
\end{aligned}
\end{equation}

To examine the wave-packet evolution at a fixed incident energy across different barrier heights, the same incident wave packet is used while varying the barrier height over a series of values. The wave packet has a central momentum $k_0 = 1.25$, corresponding to a kinetic energy of $E_k = 0.5$, and a spatial width $\sigma = 2$. The potential barrier parameters are set as $\epsilon = 10$, $a = 10$, and $d = 16.8$ (in natural units). A spatial grid of length $l = 1000$ with $N_x = 2^{14}$ points and a time step $\delta t = 10^{-3}$ is employed for the numerical simulations.
 
Fig.~\ref{wavepackets} demonstrates the time evolution process in which the efficiency of transmission can be directly observed. Apparently, the transmission efficiency in the above-barrier zone and in the normal Klein zone (above the supercritical threshold) is significantly higher than that in the normal-quantum-tunneling zone and in the special Klein zone (below the supercritical threshold). As illustrated in Fig.~\ref{perfect tunneling}, the perfect-transmission curves in the above-barrier zone and in the supercritical Klein zone are more densely distributed. A denser distribution of perfect-transmission curves implies that more components of the wave packet can transmit, resulting in higher transmission efficiency,  which is consistent with the results shown in Fig.~\ref{wavepackets}. In addition, the wave-packet remains free of exponential decay in the whole Klein zone, different from that in the normal-quantum-tunneling zone where the wave packet decays rapidly.

Moreover, enhancement of the wave packet near the barrier edge is observed in the Klein zone above the supercritical threshold as shown in Fig.~\ref{wavepackets}d, which signals that particle-antiparticle pair creation happens in this region\cite{Alkhateeb2021}. In contrast, no pronounced enhancement is observed in Fig.~\ref{wavepackets}(c), that is to say, particle-antiparticle pair does not manifest in the Klein zone below the supercritical threshold. These observations suggest that, although both areas belong to the Klein zone, the underlying physical mechanism on the two sides of the supercritical threshold may differ. Such understanding coincides with above analysis in Sec.~\ref{sec:level3}B.

\section{\label{sec:level6}Conclusion} 
In this work, we have derived the conditions for perfect transmission of a Dirac particle through a one-dimensional double barrier. It is demonstrate that a finite distance between the barriers enables perfect transmission in the normal-quantum-tunneling zone, where the transmission probability typically decays exponentially. The perfect transmission curve in the normal-quantum-tunneling zone connects continuously with the curves in both the adjacent Klein and above-barrier zones. In addition, we find that, perfect transmission can occur when the barrier height exceeds $2mc^2$ but remains below the supercritical threshold, and which appears not to require an involvement of spontaneous particle–antiparticle pair creation. These findings link two zones often considered to have fundamentally different physical properties, suggesting an underlying correlation between PKT and RT. We have also investigated time evolution of a wave-packet propagating across a double barrier with different heights, corresponding to tunneling in different energy zones. The transmission efficiency and decay behavior of the wave packet exhibit distinct features in different zones, which can be partially explained according to the distribution of perfect transmission curves in these zones.

Additionally, we discussed interesting special cases of perfect transmission in this paper, namely the zero-energy resonance and the zero-momentum resonance conditions. Beyond the fact that only the even-parity part of their conditions is influenced by the distance, we found that as the distance $d$ increases, a decrease in the minimum well depth for zero-energy resonance and an increase in the minimum barrier height for zero-momentum resonance are observed. We also derived the bound state solutions for the double-well model. Our results confirm that the supercritical state and the half-bound state conditions correspond to the zero-momentum resonance and the zero-energy resonance conditions, respectively.

\textbf{Acknowledgments} This work is supported by the National Natural Science Foundation of China (Grant No. 11874083).

\bibliographystyle{unsrt}

\begin{thebibliography}{99}
\bibitem{Bohm1951}D. Bohm, Quantum Theory (Prentice-Hall, New York, 1951).
\bibitem{Merzbacher1998}E. Merzbacher, QuantumMechanics, 3rd ed. (Wiley, NewYork,
 1998).
\bibitem{Dombey1999}N. Dombey and A. Calogeracos, Seventy years of the klein paradox. Phys. Rep. 315, 41(1999).
\bibitem{Klein1929}O. Klein, Die reflexion von elektronen an einem potentialsprung nach der relativistischen dynamik von Dirac. Z. Phys. 53, 157(1929).
\bibitem{Novoselov2004}K. Novoselov, A. Geim, S. Morozov, D. Jiang, Y. Zhang, S. Dubonos, I. Grigorieva, and A. Firsov, Electric field effect in atomically thin carbon films, Science 306, 666 (2004).
\bibitem{Castro}A. H. Castro Neto, F. Guinea, N. M. R. Peres, K. S. Novoselov, and A. K. Geim, The electronic properties of graphene, Rev. Mod. Phys. 81, 109 (2009).
\bibitem{Geim2006} M. Katsnelson, K. Novoselov, and A. Geim, Chiral tunnelling
 and the Klein paradox in graphene, Nat. Phys. 2, 620 (2006).
\bibitem{Kim2009}A. F. Young and P. Kim, Quantum interference and Klein tunnelling in graphene heterojunctions, Nat. Phys. 5, 222 (2009).
\bibitem{Bahat2010} O. Bahat-Treidel, O. Peleg, M. Grobman, N. Shapira, M. Segev, and T. Pereg-Barnea, Klein Tunneling in Deformed Honeycomb Lattices, Phys.Rev.Lett.104, 063901 (2010).
\bibitem{Illes2017}E. Illes and E. J. Nicol, Klein tunneling in the $\alpha-T_3 $ model, Phys. Rev. B 95, 235432 (2017).
\bibitem{Du2018} R. Du, M.-H. Liu, J. Mohrmann, F. Wu, R. Krupke, H. von L{\"o}hneysen, K. Richter, and R. Danneau, Tuning anti-Klein to Klein tunneling in bilayer graphene, Phys. Rev. Lett. 121, 127706 (2018).
\bibitem{Wang2022}J. Wang and J-F. Liu, Super-Klein tunneling and electronbeam collimation in the honeycomb superlattice, Phys. Rev. B 105, 035402 (2022).
\bibitem{Elahi2024}M. M. Elahi, H. Vakili, Y. Zeng, C. R. Dean, and A. W. Ghosh, Direct evidence of Klein and anti-Klein tunneling of graphitic electrons in a Corbino geometry, Phys. Rev. Lett. 132, 146302 (2024).
\bibitem{Dreisow2012}F. Dreisow, R. Keil, A. Tünnermann, S. Nolte, S. Longhi, and A. Szameit, Klein tunneling of light in waveguide superlattices, Europhys. Lett. 97, 10008 (2012).
\bibitem{Zhang2022}Z. Zhang, Y. Feng, F. Li, S. Koniakhin, C. Li, F. Liu, Y. Zhang, M. Xiao, G. Malpuech, and D. Solnyshkov, Angular-dependent Klein tunneling in photonic graphene, Phys. Rev. Lett. 129, 233901 (2022).
\bibitem{Jiang2020}X. Jiang, C. Z. Shi, Z. L. Li, S. Q. Wang, Y. Wang, S. Yang, S. G. Louie, and X. Zhang, Direct observation of Klein tunneling in phononic crystals, Science 370, 1447 (2020).
\bibitem{Sirota2022}L. Sirota, Klein-like tunneling of sound via negative index metamaterials, Phys. Rev. Appl. 18, 014057 (2022).
\bibitem{Wu2024}H. Wu, H. He, Z. Dong, L. Ye, W. Deng, M. Ke, and Z. Liu, Super Klein tunneling in phononic Lieb lattices,Phys. Rev. Appl.21,034026(2024).
\bibitem{Salger2011}C. Bick, M. Timme, D. Paulikat, D. Rathlev, and P. Ashwin, Klein tunneling of a quasirelativistic Bose-Einstein condensate in an optical lattice, Phys. Rev. Lett. 107, 244101 (2011). 
\bibitem{Zhang2012} D.-W. Zhang, Z.-Y. Xue, H. Yan, Z. D. Wang, and Shi-Liang
 Zhu, Macroscopic Klein tunneling in spin-orbit-coupled Bose Einstein condensates, Phys.Rev.A85, 013628 (2012).
\bibitem{Greiner}W. Greiner, Relativistic Quantum Mechanics, 3rd ed. (Springer, Berlin, 1996).
\bibitem{Su1993}R. K. Su,  G. C. Siu, and X. Chou, Barrier penetration and Klein paradox. J. Phys. A 26, 1001(1993).
\bibitem{Grobe2004}P. Krekora, Q. Su, and R. Grobe, Klein paradox in spatial and temporal resolution. Phys. Rev. Lett. 92, 040406 (2004).
\bibitem{Matzkin2022}M. Alkhateeb and A. Matzkin, Space-time-resolved quantum field approach to Klein-tunneling dynamics across a finite barrier, Phys.Rev.A106, L060202 (2022).
\bibitem{Hosack1965}H. H. Hosack, Double barrier transmission characteristics, J. Appl. Phys. 36, 1281 (1965).
\bibitem{Hauge1987}E. H. Hauge, J. P. Falck, and T. A. Fjeldly, Transmission and reflection times for scattering of wave packets off tunneling barriers, Phys. Rev. B 36, 4203 (1987).
\bibitem{Lunardi2007}J. T. Lunardi and L. A. Manzoni, Relativistic tunneling through two successive barriers, Phys. Rev. A 76, 042111 (2007).
\bibitem{Bai2007}C. Bai and X. Zhang, Klein paradox and resonant tunneling in a graphene superlattice, Phys. Rev. B 76, 075430 (2007).
\bibitem{Villalba2010}V. M. Villalba and L. A. Gonz{\'a}lez-{\'A}rraga. Tunneling and transmission resonances of a dirac particle by a double barrier, Phys. Scr. 81, 025010 (2010).
\bibitem{Vargas2012}I. Rodriguez-Vargas, J. Madrigal-Melchor, and O. Oubram, Resonant tunneling through double barrier graphene systems: A comparative study of Klein and non-Klein tunneling structures Available to Purchase, J. Appl. Phys. 112, 073711 (2012).
\bibitem{Dombey2000}N. Dombey, P. Kennedy, and A. Calogeracos. Supercriticality and transmission resonances in the dirac equation. Phys. Rev. Lett. 85,  1787(2000).
\bibitem{Dombey2002}P. Kennedy, and N. Dombey. Low momentum scattering in the Dirac equation. J. Phys. A. 35, 6645(2002).
\bibitem{KennedyW}P. Kennedy, The Woods–Saxon potential in the Dirac equation, J. Phys. A: Math. Gen. 35, 689 (2002).
\bibitem{Alkhateeb2021}M. Alkhateeb, X. G. de la Cal, M. Pons, D. Sokolovski, and A. Matzkin, Relativistic time-dependent quantum dynamics across supercritical barriers for Klein-Gordon and Dirac particles, Phys. Rev. A 103, 042203 (2021).

\end{thebibliography}

\end{document}